\documentstyle[11pt]{article}

\title{ Mirror fermions and LEP precision data }

\vspace{1cm}

\author{ C. Cs\'aki and
 F. Csikor \\
 Institute for Theoretical Physics, E\"otv\"os University,
 Budapest }

\date{February, 1993}
\newcommand{\be}{\begin{equation}}
\newcommand{\ee}{\end{equation}}
\newcommand{\bd}{\begin{displaymath}}
\newcommand{\ed}{\end{displaymath}}
\newcommand{\bea}{\begin{eqnarray}}
\newcommand{\eea}{\end{eqnarray}}

\begin{document}
\maketitle

\begin{abstract} \normalsize

The three generation mirror fermion model is compared with LEP precision
and low energy data. While for zero mixing of ordinary and mirror fermions
the model is not favoured, for non zero mixing very low $\chi ^2 $ fits
have been obtained, in particular when right handed leptonic mixings are small
and left handed leptonic mixings are large (of the order of 0.1 radians.)

\end{abstract}


The validity of the standard model (SM) has been successfully tested at the one
loop level by the LEP data \cite{b1}. The precise experimental data
severly limit
the possibility of any kind of new physics. It is thus compulsory to confront
any hypothetical extension of the standard model with the precise LEP data. In
this note we consider the possible existence of mirror fermions. The very
simple extension of the SM of \cite{b2} is to enlarge only the fermion content
by introducing mirror (i.e. opposite chirality property)
fermions to {\em each}
fermion of the SM (i.e. to each ordinary fermion), preserving the
$SU(2)\otimes U(1)$  group structure. Ordinary and mirror fermions are allowed
to mix. In fact mixing is necessary in order to avoid stable mirror fermions.
Present experiments directly exclude mirror fermions with masses below roughly
half of the $Z^0$ mass. More massive mirror fermions are still allowed. In the
following we consider heavy mirror fermions with masses exceeding the weak
vector boson masses. Many of the  phenomenological consequences of such a
model have been worked out in \cite{b3}, \cite{b4}, where motivations for the
model are also discussed.

To compare the  mirror model with LEP data one has to calculate the one loop
corrections to the physical observables. This is particularly easy in the case
of no mixing between ordinary and mirror fermions. Since LEP observables refer
only to ordinary fermions it is easy to see that the effects of mirror fermions
in the loops are the same as those of heavy sequential fermions. This kind of
new physics has been considered in the literature (e.g. in \cite{b5},
\cite{b6}, \cite{A},
\cite{b7}). The modification of the SM predictions can be expressed in terms of
the familiar S, T, U (or the equivalent $\epsilon _i $ ) variables. The heavy
fermions modify only the vector boson self energies. The latest analysis by
Ellis et al. \cite{b7} (including LEP data, $W$ mass and low energy data)
yields:
$\tilde{\epsilon _1} =(0.09\pm 0.25)\times 10^{-2} $ ($\tilde{T} =0.01233\pm
0.3425 $)
$\tilde{\epsilon _2} =(0.09\pm 0.72)\times 10^{-2} $ ($\tilde{U} =-0.1125\pm
0.90 $)
$\tilde{\epsilon _3} =-(0.24\pm 0.38)\times 10^{-2} $ ($\tilde{S} =-0.3024\pm
0.48 $),
where the tilde refers to the deviation of the variables from the SM reference
point (i.e. $m_{top} =130 GeV$, $M_{Higgs} =M_Z $.) The nonzero values may
represent the new physics. The contribution of a heavy fermion family
consisting of degenerate doublets to T and U is vanishing, while S gets a
positive contribution of $2/3\pi $. The possibility of new heavy (ordinary or
mirror) unmixed fermion generations is thus severely constrained.
(The 2$\sigma $ upper bound on $\tilde{S} $ is only slightly larger
than the mirror contribution.)

Though the above scenario is quite discouraging, it does not yet exclude the
possibility of three heavy generations, because of the possibility of mixing.
The mixing schemes of sequential and mirror fermions are different. In the
following we are only concerned with the possibility of mirror fermions. The
obvious modification to the case of zero mixing is that tree level couplings
change. As a consequence {\em one loop corrections to LEP observables
can not be
discussed in the usual way} considering the experimental determinations of
$\tilde{S} $, $\tilde{T}$, $\tilde{U}$.

The general mixing scheme in the mirror fermion model has been considered in
\cite{b8}. It is possible that left and right handed fields mix with different
mixing angles. Denoting mixing angles by $\alpha ^f _L$ and $\alpha ^f _R$
the tree level couplings are easily obtained from the appropriate part of the
Lagrangean. For the neutrino - electron (ordinary and mirror) doublets
e.g. we have

\begin{eqnarray}
\lefteqn{{\cal L}_{CC} = }\nonumber  \\
 & & - \frac{e_0}{2 \sqrt{2} s2} \left[ \ ( \bar{e}
\cos \alpha _L ^e +\bar{E} \sin \alpha _L ^e ) \gamma _{\mu} (1-\gamma _5 )
(\nu \cos \alpha _L ^\nu + N \sin \alpha_L ^\nu )\ +\right. \nonumber \\
& & \left. ( \ -\bar{e}
\sin \alpha _R ^e +\bar{E} \cos \alpha _R ^e ) \gamma _{\mu} (1+\gamma _5 )
(-\nu \sin \alpha _R ^\nu + N \cos \alpha_R ^\nu )\ \right]
\cdot W^{\mu}+ \nonumber \\  & & {\rm h.c.}
\end{eqnarray}
\newpage
\begin{eqnarray}
\lefteqn{{\cal L}_{NC} = -\frac{e_0}{2s2 c2} }\nonumber \\
 & &   \left[ \ ( \bar{e} \right.
\cos \alpha _L ^e +\bar{E} \sin \alpha _L ^e ) \gamma _{\mu} (-\frac{1}{2}+s2)
(1-\gamma_5)(e \cos \alpha _L ^e+ E \sin \alpha_L ^e )\ + \nonumber \\
& & (-\bar{e} \sin \alpha _R ^e +\bar{E} \cos \alpha _R ^e )
\gamma _{\mu} (-\frac{1}{2}+s2)
(1+\gamma_5)(-e \sin \alpha _R ^e+ E \cos \alpha_R ^e )\ + \nonumber \\
& & (-\bar{e} \sin \alpha _L ^e +\bar{E} \cos \alpha _L ^e ) \gamma _{\mu} s2
(1-\gamma_5)(-e \sin \alpha _L ^e+ E \cos \alpha_L ^e )\ + \nonumber \\
& & \left. (\bar{e} \cos \alpha _R ^e +\bar{E} \sin \alpha _R ^e )
\gamma _{\mu} s2
(1+\gamma_5)(e \cos \alpha _R ^e+ E \sin \alpha_R ^e )\ \right]
\cdot Z^{\mu} \nonumber \\
& & -\frac{e_0}{4 s2 c_W} \left[  \ ( \bar{\nu }
\cos \alpha _L ^\nu  +\bar{N} \sin \alpha _L ^\nu  ) \gamma _{\mu}
(1-\gamma_5)(\nu  \cos \alpha _L ^\nu + N \sin \alpha_L ^\nu  )\ +\right.
\nonumber \\
& & \left. (-\bar{\nu} \sin \alpha _R ^\nu  +\bar{N} \cos \alpha _R ^\nu  )
\gamma _{\mu}
(1+\gamma_5)(-\nu  \sin \alpha _R ^\nu + N \cos \alpha_R ^\nu  )\ \right]
\cdot Z^{\mu} \nonumber \\
& & -e_0 j_\mu ^{el} \cdot A^\mu .
\end{eqnarray}

\noindent where $e_0$ is the proton charge, the $\alpha $'s  denote the
mixing angles and $s2=1-c2 \equiv \sin^2
\theta_W$. The interactions for other fermion doublets have similar forms with
different   mixing angles. Note that we have excluded possible
intergeneration mixings and CP violation.

Due to the special structure of mass matrices (mirror masses are generated
by spontaneous symmetry breaking at the electroweak scale, ordinary-mirror
mixing is a consequence of direct coupling of the 'current' fields), there is
an unusual relation among the masses and mixing angles of a particular
doublet (ordinary and mirror) (\cite{b8}). For the neutrino - electron doublets
e.g. we have:

\begin{eqnarray}
\lefteqn{-m_{\nu} \sin \alpha^\nu _R \cos \alpha^\nu _L +m_N
\cos \alpha^\nu _R \sin \alpha^\nu _L \,=\, } \nonumber \\
 & & -m_{e} \sin \alpha^e _R \cos \alpha^e _L +m_E \cos \alpha^e _R \sin
\alpha^e _L.
\end{eqnarray}

\noindent Such relations are very important in the calculation of loop
corrections, since they play a role in the cancellation of the divergencies.

Experimental information on the mixing angles (in the form of upper bounds) is
given in \cite{b9} and \cite{b10}. Mixing angles are bounded typically by
0.1 - 0.2.
A different and often much stronger bound is obtained for leptonic mixing
angles using the
experimental constraint on the mirror fermion contribution to the anomalous
magnetic moments of the electron and the muon. E.g. assuming equal
mixing angles in a lepton doublet the bound is 0.02 (as given in \cite{b8}.)
These small mixing angles result in too small cross-sections, so that mirror
fermion production at HERA would be undetectable. As discussed in \cite{b8}
HERA
cross-sections may be saved assuming $\alpha _R ^{e} \approx 0$, ensuring small
anomalous magnetic moment contributions without restricting $\alpha ^{e,\nu} _L
$ and $\alpha _{R} ^{\nu } $.

Using the couplings of Eqs. (1), (2) for all the doublets  it is
straightforward to
calculate the $Z^0 \rightarrow f\bar{f}$ effective couplings at the scale of
the $Z^0$
mass. We have used the on shell scheme. New diagrams (as compared to the SM
case) occur in the vector boson propagator corrections, vertex and box
corrections. The $G_\mu$-$M_W $ relationship and the running of $\alpha $ also
changes. A precise calculation of all the diagrams is quite lengthy, though
straightforward. However, since we are concerned with the effects of heavy
fermions, the dominant effect comes from vector boson propagator effects. We
have therefore calculated these corrections precisely and approximated the rest
(i.e. vertex and box corrections) by the zero mixing angle formulae. Our
results for the effective couplings are

\begin{eqnarray}
v_f & = & \sqrt{\rho _f }\, (t_{3f} \,({\cos}^2 \alpha _L ^f +
{\sin}^2 \alpha _R ^f )
  -2Q_f {\sin}^2 \overline {\Theta _W }) \\
a_f & = & \sqrt{\rho _f }\, t_{3f} \, ({\cos}^2 \alpha _L ^f
-{\sin}^2 \alpha _R ^f ) \\
\rho _f & = & \rho ^{SM} +\alpha \tilde{T} + 2{\epsilon }^2 \\
{\sin}^2 \overline {\Theta _W } & = & {{\sin}^2 \overline {\Theta _W }}^{SM}
-\frac{2 c^2 s^2 }{c^2 -s^2 }{\epsilon }^2
+\frac{\alpha }{4(c^2-s^2)}
(\tilde {S} -4c^2 s^2  \tilde {T} ),
\end{eqnarray}
\noindent where the terms containing $\epsilon $ arise from the modification
of $G_\mu $ in the mirror model. We have

\begin{eqnarray}
{\epsilon }^2 =\frac{1}{2}\left[ 1-\left( \frac{(1+a^2 )^2 +4a^2
}{8}\right) ^{\frac{1}{2}}
(\cos \alpha _L ^l \cos \alpha _L ^\nu
+\sin \alpha _R ^l \sin \alpha _R ^\nu )^2 \right],
\end{eqnarray}
\noindent where lepton universality (i.e. $\alpha ^\mu =\alpha ^e $,
$\alpha ^{\nu _{\mu}}
 =\alpha ^{\nu _e} $) and the smallness of mixing angles (so that the 4th
powers are negligible) has been assumed.

The $M_W $ - $M_Z $ relationship changes to

\begin{eqnarray}
{\cos }^2 \Theta _W ^{SM}  \rightarrow {\cos }^2 \Theta _W ^{SM} (1+
\frac{2s^2 }{c^2 -s^2 }{\epsilon }^2 +
\nonumber \\
\frac{\alpha }{4(c^2 -s^2 )s^2 }
(4c^2 s^2 \tilde {T} -2s^2 \tilde {S} +(c^2 -s^2 )\tilde {U} )),
\end{eqnarray}

\noindent where ${\cos }^2 \Theta _W ^{SM} $ is the usual SM expression
(in terms of
$\alpha $, $G_\mu $ and $M_Z $.)

Evaluation of the SM correction has been
performed using the ZFITTER - DIZET package of \cite{b11}.
Note that the ordinary fermion contributions to $S$, $T$, $U$ are not finite
for nonzero mixing. Cancellation of divergencies is achieved only after
including  mirror fermion and mixed ordinary - mirror fermion
loops. Therefore we have directly determined the complete $S$,
$T$ and  $U$ for non-zero mixing, subtracting the standard model
(i.e. zero mixing, mirrors excluded) contribution.
Eq. (3) and similar relations play a decisive role in the cancellation of
divergencies.

In the mirror fermion model we have a large number of free parameters. For
each doublet there are 4 mixing angles and the two masses of the mirror
partners. Taking into account Eq. (3) we have 5 parameters.
To make a meaningful comparison with experimental data the number of free
parameters should be limited by assuming different physical scenarios,
specifying the various parameters \cite{b13}. Since for zero mixing and
large mirror
masses (larger than 100 GeV) three mirror generations are already excluded by
LEP data, it is satisfactory if consistency of (the three generation) mirror
fermion model can be demonstrated for acceptable  non zero mixing angle and
mass values.  Though the number of parameters is large it is by no means
trivial that such parameters do indeed exist.
It is reasonable to start with degenerate mirror doublets, so that $\tilde{T} $
and $\tilde{U}$ are small. The reasonable range for the mirror masses is from
100GeV to 500 GeV. The upper limit is suggested by the well known perturbative
tree unitarity mass bounds worked out in \cite{b4} for the mirror model.
These bounds are
particularly strong for the case of degenerate multiplets.  The following cases
have been studied \\
I. $\alpha _L ^l =\alpha _R ^l =\alpha _L ^\nu =\alpha _R ^\nu$;
$\alpha _L ^q =\alpha _R ^q$ \\
II. $\alpha _R ^\nu =0$, $\alpha _L ^l =\alpha _R ^l  =\alpha _L ^\nu$;
$\alpha _L ^q =\alpha _R ^q$ \\
III. $\alpha _R ^l =\alpha _R ^\nu =0$, $\alpha _L ^l =\alpha _L ^\nu$;
$\alpha _L ^q =\alpha _R ^q$ , \\
where $\alpha ^l$ refers to charged leptons, $\alpha ^\nu$ to neutrinos and
$\alpha ^q $ to u,d,c or s quarks. Assuming lepton universality we take equal
leptonic mixing angles for the three generations. The top -bottom doublet
clearly plays a special role. Therefore top and bottom mixing angles are taken
different from other quark's mixing angles, while the latter are assumed to be
equal. Also masses of mirror top and mirror bottom should be larger than other
mirror masses and in any case larger than the top mass. We take
$M_{mirror}=$100 GeV (200 GeV) for the isospin 1/2 mirrors except that
$M_{m.top} =M_{m.bottom} =$200 GeV (300GeV). Masses of isospin -1/2 mirrors
(except for m.bottom) and $\alpha _L ^{top} =\alpha _R ^{top} $ are determined
from Eq. (3). It turns out that our fits are very insensitive to the actual
values of masses (the cases of the lower and higher masses are practically
indistinguishable), the important assumption
\newpage \vspace*{-0.5cm} \noindent
is degeneracy.

Case I has a very simple meaning, namely at tree level only the axial vector
parts of the  weak currents are modified by a mixing angle factor. Case II
differs only slightly from case I, namely  mixing of right handed neutrinos
is not allowed (therefore  ordinary neutrinos are purely left handed.)
In these two cases leptonic mixing angles are small (less than 0.02) due to the
constraints arising from the anomalous magnetic moments of electron and muon
as discussed in \cite{b8}. Case III is not restricted by these limits (the
mirror fermion contributions to the  anomalous magnetic moments being zero),
therefore the leptonic mixing angles may be larger than 0.02 and are bounded
from above by $\approx $0.1 \cite{b9}, \cite{b10}.

The LEP data (and W mass) we used in the fit are given in \cite{b1} together
with the correlation matrix given in \cite{b12}.
We use $\Gamma _{total}$,
 ${g_v ^e}^2 $,  ${g_a ^e}^2 $, $A_{pol} ^{\tau } $, $\sigma _{hadron}
^{pole} $,  $A_{FB} ^{bb} $ and $1-M_W ^2 /M_Z ^2 $. We could not find a clear
interpretation of the $q\bar{q} $ charge asymmetry in the mirror fermion model,
so we excluded it from the fit. For comparison we quote \cite{b1} which gives
for the standard model fit (case A) $\chi ^2 $/d.o.f.=2.6/5,  for the central
values $m_{top} =139 $ GeV, $M_ {Higgs} =300$ GeV and $\alpha _s =$0.135. In
our fits we fix the leptonic mixing angles, the mirror masses (except
for those which are determined by Eq. (3)), top and Higgs masses as well as
$\alpha _s $. Thus we fit the quark and bottom quark mixing angles only, with
the number of degrees of freedom equal to 5. Our main concern is to see whether
or not nonzero mixing angles allow for a better fit for the mirror model than
for zero mixing angles. Fig. 1 shows the equal $\chi ^2 $ curves for case I
and the fixed parameter values given in the figure caption. The
minimum $\chi ^2 $ is 6.125 (minimized with respect to the top
and Higgs masses too.)
For the same parameters and zero mixing angle $\chi ^2 $=8., so the fit is
worse. We do not show a figure for case II since it is very close numerically
to case I. Fig. 2 shows the equal $\chi ^2 $ curves  for case III and the
fixed parameters given in the figure caption. The
minimum $\chi ^2 $ is 1.95 (minimized with respect to the top
and Higgs masses too.) It is interesting that $\alpha _q$=0 is excluded
from the very low $\chi^2 $ region.
For the same parameters but zero mixing angles $\chi ^2 $=14.65 , so our fit
gives a dramatic improvement.
 We see that acceptable fits have been obtained for all the cases, while
case III gives a very good fit. The range of parameters giving
comparable fits in case III is: $\alpha ^l _L \in $(0.08 - 0.11),
$m_{top} \in $(100,160) GeV, $M_{Higgs} \in $(70,500). The
favoured top mass is 110-120 GeV, while the Higgs mass
dependence is very small.

It is reasonable to add low neutrino scattering and atomic
parity violation
data to the fit, since they have a nonnegligible effect \cite{b13}.
\newpage \vspace*{-0.5cm}
\noindent
The results
of a model independent determination of the couplings  are given in
\cite{b14}.  We have used all the nine observables determined experimentally,
i.e.  $g_L ^2 $, $g_R ^2$, $\theta _L $, $\theta _R $, $g_A ^e $,
$g_v ^e $, $C_{1u} $, $C_{1d} $ and $C_{2u} -\frac{1}{2} C_{2d} $ together
with the correlation matrix given in \cite{b14}.
In the fit it is assumed
that neutrino couplings are pure left handed so we may compare the mirror
model with data only in case II and III.

To estimate the one loop corrections
we have made the same approximation as for the LEP case, i.e. included
nonzero mixing
angles at the tree level as well as in the vector boson propagators.
The equal $\chi ^2$ curves for
cases II, III are shown in Figs. 3 and 4.  The no. of
degrees of freedom of the fits is 14. The fits are good for essentially the
same
parameters and the
confidence levels are
increased as compared to the fits of only LEP (and $M_W $) data.

All the above fits have been made assuming degenerate mirror doublets. It is an
interesting question to ask how much this condition can be relaxed.
We have found that a 50 GeV mass splitting in the mirror top -
bottom doublet does not spoil the good fit of case III, a 100
GeV splitting is still acceptable ($\Delta \chi ^2 \approx 4$), while
150 GeV splitting is excluded. The surprisingly small sensitivity to the
doublet mass splitting is due to Eq. (3), which correlates the
mass changes with an appropriate change in the mixing angles.

In conclusion the mirror fermion model with three heavy mirror generations
and {\em no mixing} is
ruled out by LEP data at the 90\% confidence level. Allowing for nonzero mixing
the model is still alive. Since nonzero mixing is necessary in order to allow
for decay of mirror particles, we think that the mirror model has survived the
test of LEP data. Due to the large number of parameters a determination of the
model parameters is not possible. We have found particular values of the model
parameters which allow for a favourable comparison of the model predictions
with data. In particular zero right leptonic mixing angles and
large (of the order of 0.1 radians) left leptonic mixing angles lead to
excellent fits of experimental data.

\vspace{1cm}

\large\bf Acknowledgements \normalsize\rm \newline

\vspace{1cm}

We thank G. Fogli and D.Schildknecht for  discussions. F. Cs. thanks
M. Bilenky for advice  on how to use data and help with interpretation
of ZFITTER output.

\newpage
\vspace{2cm}
\begin{center}  \Large\bf Figure captions \normalsize\rm
\end{center}
\vspace{1cm}

\bf Fig.\,1. \hspace{5pt} \rm
Equal $\chi ^2$ curves of a fit to LEP data. The fixed parameters are
$m_{top} =$150 GeV, $M_{Higgs} =$70 GeV, $\alpha _s =$0.14,
$\alpha _L ^l =\alpha _R ^l =\alpha _L ^\nu =\alpha _R ^\nu =$0.02.
$\alpha _L ^q =\alpha _R ^q$,  $\alpha _L ^b =\alpha  _R ^b $ and
$\alpha _L ^{top} =\alpha  _R ^{top} $.
The area below the  indicated curve is the lowest  $\chi ^2$ area. The
next curves correspond to $\chi ^2$'s increasing by steps of 0.5.
\vspace{10pt}

\bf Fig.\,2. \hspace{5pt} \rm
Equal $\chi ^2$ curves of a fit to LEP data. The fixed parameters are
$m_{top} =$110 GeV, $M_{Higgs} =$200 GeV, $\alpha _s =$0.14,
$\alpha _R ^l =\alpha _R ^\nu =$0.0,
$\alpha _L ^l =\alpha _L ^\nu  =$0.09.
$\alpha _L ^q =\alpha _R ^q$,  $\alpha _L ^b =\alpha  _R ^b $ and
$\alpha _L ^{top} =\alpha  _R ^{top} $.
The area between  the  indicated curves is the lowest  $\chi ^2$ area. The
next curves correspond to $\chi ^2$'s increasing by steps of 0.5.
\vspace{10pt}

\bf Fig.\,3. \hspace{5pt} \rm
Equal $\chi ^2$ curves of a fit to LEP and low energy data. The fixed
parameters are
$m_{top} =$150 GeV, $M_{Higgs} =$70 GeV, $\alpha _s =$0.14,
$\alpha _R ^\nu =$0,
$\alpha _L ^l =\alpha _R ^l =\alpha _L ^\nu =
$0.02.
$\alpha _L ^q =\alpha _R ^q$,  $\alpha _L ^b =\alpha  _R ^b $ and
$\alpha _L ^{top} =\alpha  _R ^{top} $.
The area below the  indicated curve is the lowest  $\chi ^2$ area. The
next curves correspond to $\chi ^2$'s increasing by steps of 0.5.
\vspace{10pt}

\bf Fig.\,4. \hspace{5pt} \rm
Equal $\chi ^2$ curves of a fit to LEP and low energy data. The fixed
parameters are
$m_{top} =$110 GeV, $M_{Higgs} =$300 GeV, $\alpha _s =$0.14,
$\alpha _R ^l =\alpha _R ^\nu =$0.0,
$\alpha _L ^l =\alpha _L ^\nu  =$0.09.
$\alpha _L ^q =\alpha _R ^q$,  $\alpha _L ^b =\alpha  _R ^b $ and
$\alpha _L ^{top} =\alpha  _R ^{top} $.
The area between the  indicated curve is the lowest  $\chi ^2$ area. The
next curves correspond to $\chi ^2$'s increasing by steps of 0.5.

\end{document}